\documentclass[sigconf,nonacm]{acmart}

\AtBeginDocument{%
  \providecommand\BibTeX{{%
    \normalfont B\kern-0.5em{\scshape i\kern-0.25em b}\kern-0.8em\TeX}}}

\usepackage{multirow} 
\usepackage{tabularx}
\usepackage{makecell}
\usepackage{hyperref}

\graphicspath{{./images/}}

\begin{document}


\title{TTT4Rec: A Test-Time Training Approach for Rapid Adaption in Sequential Recommendation}

\author{Zhaoqi Yang}
\email{zyang3@arizona.edu}

\affiliation{%
  \institution{MIS Department, Eller College of Management, University of Arizona}
  \city{Tucson}
  \state{Arizona}
  \country{USA}
}

\author{Yanan Wang}
\email{ynwwang@arizona.edu}

\affiliation{%
  \institution{MIS Department, Eller College of Management, University of Arizona}
  \city{Tucson}
  \state{Arizona}
  \country{USA}
}

\author{Yong Ge}
\email{yongge@arizona.edu}

\affiliation{%
  \institution{MIS Department, Eller College of Management, University of Arizona}
  \city{Tucson}
  \state{Arizona}
  \country{USA}
}

\renewcommand{\shortauthors}{Yang, et al.}

\begin{abstract}
Sequential recommendation tasks, which aim to predict the next item a user will interact with, typically rely on models trained solely on historical data. However, in real-world scenarios, user behavior can fluctuate in the long interaction sequences, and training data may be limited to model this dynamics. To address this, Test-Time Training (TTT) offers a novel approach by using self-supervised learning during inference to dynamically update model parameters. This allows the model to adapt to new user interactions in real-time, leading to more accurate recommendations. In this paper, we propose \textbf{TTT4Rec}, a sequential recommendation framework that integrates TTT to better capture dynamic user behavior. By continuously updating model parameters during inference, TTT4Rec is particularly effective in scenarios where user interaction sequences are long, training data is limited, or user behavior is highly variable. We evaluate TTT4Rec on three widely-used recommendation datasets, demonstrating that it achieves performance on par with or exceeding state-of-the-art models. The codes are available at \href{https://github.com/ZhaoqiZachYang/TTT4Rec}{https://github.com/ZhaoqiZachYang/TTT4Rec} and the data is available at \href{https://drive.google.com/drive/folders/1ugjwxz_QNZqfuDdpzJs2GB7lDBv1Pm31?usp=sharing}{here}.

\end{abstract}


\maketitle
\vspace{-0.3cm}

\section{Introduction}
Sequential recommendation systems aim to predict the next item a user will interact with by modeling the dependencies within their interaction history. Over the years, a variety of deep learning models have been developed to capture these sequential patterns. Early approaches employed Recurrent Neural Networks (RNNs) \cite{hidasi2015session,li2017neural}, which utilize hidden states to store historical information. However, RNNs often suffer from issues like vanishing gradients and limited capacity for long-term dependencies. More recently, attention-based models like SASRec \cite{kang2018self} and BERT4Rec \cite{sun2019bert4rec} have emerged as strong alternatives due to their superior performance in modeling complex user behavior. Despite their success, these models come with significant computational costs, especially when handling long interaction sequences, due to the quadratic complexity of self-attention mechanisms \cite{tay2020long}. In contrast, recent advances in state-space models (SSMs) \cite{gu2021efficiently}, like Mamba4Rec \cite{liu2024mamba4rec}, have demonstrated a capacity to model long-range dependencies with linear complexity. Nonetheless, one major limitation persists: these models are static at deployment, relying solely on the information learned during training and lacking the ability to adapt to evolving user behavior once deployed.

To overcome this limitation, Test-Time Training (TTT) \cite{sun2024learning} introduces an innovative framework that allows models to update their parameters during inference. Unlike traditional methods, which freeze model parameters after training, TTT employs a self-supervised learning process that continues to refine model parameters based on user interactions during testing. This approach enables the model to adjust in real-time, improving its ability to capture dynamic patterns in user behavior. The TTT framework operates through a dual-loop structure: the \textbf{outer-loop} focuses on supervised learning during training, while the \textbf{inner-loop} employs self-supervised learning, enabling continuous updates to the model's hidden states during both training and inference. This ability to adapt to new data during test time provides significant improvements, especially in cases where user preferences shift rapidly or when training data is limited.

In this work, we introduce TTT4Rec, a novel sequential recommendation framework that incorporates TTT to enable real-time model adaptation. By continuously updating its parameters during inference, TTT4Rec dynamically responds to changing user behaviors. The architecture of TTT4Rec consists of several key components. First, an embedding layer maps item IDs to a high-dimensional space, representing user-item interactions. Positional embeddings are then applied to preserve the temporal order of these interactions. Several residual blocks are stacked to increase the depth of the model while ensuring smooth information propagation. The core component, the TTT layer, applies self-supervised learning during both training and inference to ensure that model parameters are continuously updated in response to new user interactions. Finally, the prediction layer converts the updated hidden states into a probability distribution over all items, enabling accurate next-item recommendations.

We evaluate TTT4Rec on three widely-used benchmark datasets for sequential recommendation, covering a variety of real-world scenarios, including location-based check-ins, video streaming, and online shopping. Comparison experiments show that TTT4Rec consistently outperforms state-of-the-art models by dynamically adapting to real-time user interactions. Our results highlight TTT4Rec’s effectiveness, particularly in scenarios with limited training data or volatile user behavior, where traditional models often fall short.

The rest of the paper is organized as follows: In Section 2, we present the preliminaries of sequential recommendation and the fundamentals of TTT. In Section 3, we detail the architecture of TTT4Rec and describe each of its components. Section 4 presents our experimental setup and results. Finally, Section 5 concludes with a discussion of potential future research directions.

\section{Preliminaries}

\subsection{Sequential Recommendation}
In sequential recommendation, the goal is to predict the next item that a user will interact with based on their historical interaction sequence. Formally, let $\mathcal{U} = \{u_1, u_2, \dots, u_{|\mathcal{U}|}\}$ denote the set of users, and $\mathcal{V} = \{v_1, v_2, \dots, v_{|\mathcal{V}|}\}$ denote the set of items. For each user $u \in \mathcal{U}$, the interaction sequence $S_u = \{v_1, v_2, \dots, v_{n_u}\}$ represents the ordered items that the user has interacted with, where $n_u$ is the length of the sequence. Given $S_u$, the goal of the model is to predict the next item $v_{n_u+1}$ that user $u$ will interact with.

\subsection{Test-Time Training (TTT)}
Test-Time Training (TTT) \cite{sun2024learning} introduces a new approach to sequence modeling by allowing model parameters to be updated during test time based on new information. At the core of TTT layer is the idea that the model’s hidden state is itself a learnable machine learning model. Instead of simply passing information through fixed parameters (as in RNNs or Transformers), TTT updates its hidden state parameters using gradient-based optimization. During test time, the inner-loop performs gradient updates using a self-supervised loss, enabling the model to adapt in real-time. Specifically, for a given input sequence $x_t$, the model updates its parameters as follows:
\[
W_t = W_{t-1} - \eta \nabla \mathcal{L}(W_{t-1}; x_t),
\]
where $W_t$ represents the model state at step $t$, $\eta$ is the learning rate, and $\mathcal{L}$ is the self-supervised loss.

\section{TTT4Rec}

In this section, we introduce \textbf{TTT4Rec}, a novel sequential recommendation framework that leverages Test-Time Training (TTT) to dynamically update model parameters during inference. As shown in Fig.~\ref{fig_stru}, TTT4Rec integrates three main components: an embedding layer that encodes item information into high-dimensional vectors, one or more TTT-based residual blocks that capture characteristics of the input sequence, and a prediction layer that generates recommendations based on the model's updated hidden states. 

\subsection{Embedding Layer}

The Embedding Layer in TTT4Rec maps item IDs to high-dimensional vectors, providing a dense representation of each item in the recommendation task. We use an embedding matrix \( E \in \mathbb{R}^{|V| \times D} \), where \( |V| \) is the size of the item set and \( D \) is the embedding dimension. Given a sequence of items \( S_u = \{v_1, v_2, \dots, v_{n_u}\} \) for user \( u \), the embedding layer maps each item \( v_i \) in the sequence to its corresponding vector representation \( e_i = E(v_i) \).

After obtaining the item embeddings, we apply Rotary Embedding (RoPE) \cite{su2024roformer} to encode positional information. Formally, for an item embedding \( e_i \), the rotary position embedding \( e_i^{\text{rot}} \) is computed as follows:

\[
e_i^{\text{rot}} = \text{RoPE}(e_i, i; \mu),
\]
where \( i \) is the position of the item in the sequence, \( \mu \) is a hyperparameter that adjusts the degree of a single step rotation, and RoPE applies a rotation matrix to the embedding \( e_i \), encoding its position relative to other items. The resulting position-encoded embeddings are flexible and suited for capturing sequential patterns, especially in long sequences.

To enhance the robustness of the embeddings and prevent overfitting, we apply LayerNorm and Dropout to the output of the embedding layer. The normalized and regularized embeddings are given by:

\[
H = \text{LayerNorm}(\text{Dropout}(S_u E)) \in \mathbb{R}^{n_u \times D}.
\]

In this way, the embedding layer in TTT4Rec not only captures the individual item features but also encodes the positional relationships between items, ensuring that the model can effectively represent both item-specific and sequential information in the recommendation task.

\subsection{Residual Block}
\begin{figure}[htpb]
\centering
  \includegraphics[width=0.2\textwidth]{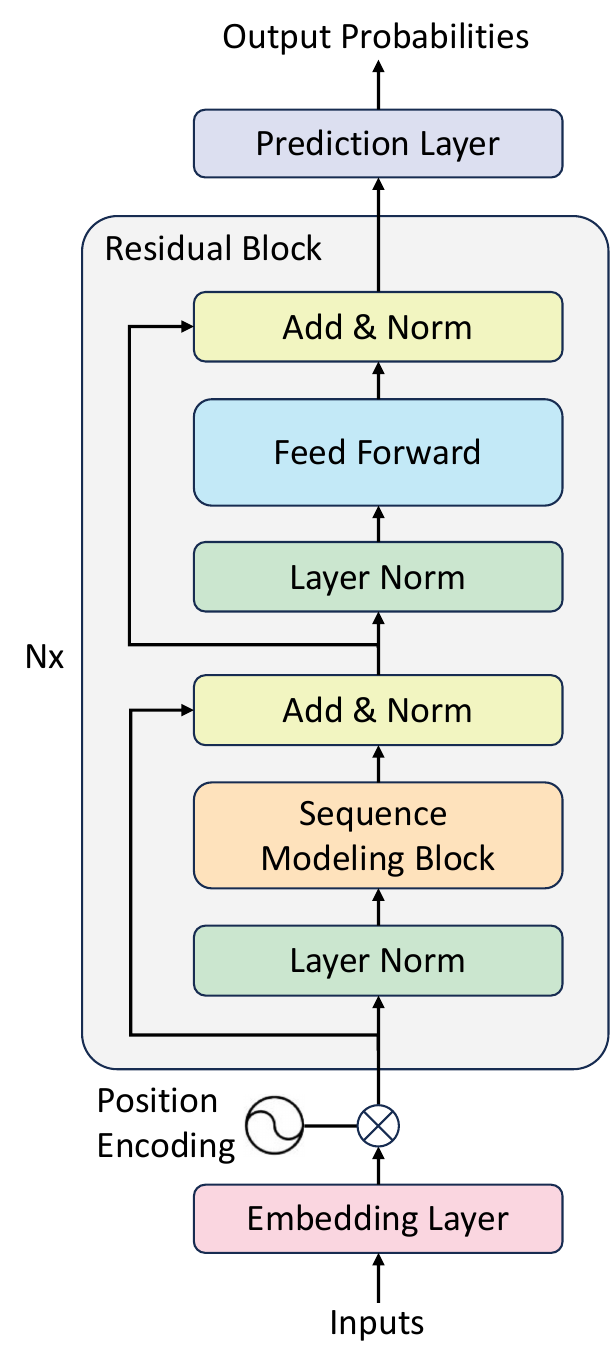}
  \caption{Structure of the TTT4Rec Model.}
  \label{fig_stru}
\end{figure}

The residual block is inspired by modern sequence modeling architectures like Transformers. As shown in Fig.~\ref{fig_stru}, each residual block consists of several key components: layer normalizations (LayerNorm), a sequence modeling block, a feed-forward block, and residual connections that span across layers to help propagate information through the network efficiently. 

Each residual block begins with a \textbf{LayerNorm} operation, which normalizes the inputs to stabilize training. Following LayerNorm is the \textbf{sequence modeling block}, which captures dependencies in the input sequence. The sequence modeling block can be instantiated into two variants: the \textbf{Transformer Backbone} and the \textbf{Mamba Backbone}. 

\begin{figure}[h]
\centering
  \includegraphics[width=0.4\textwidth]{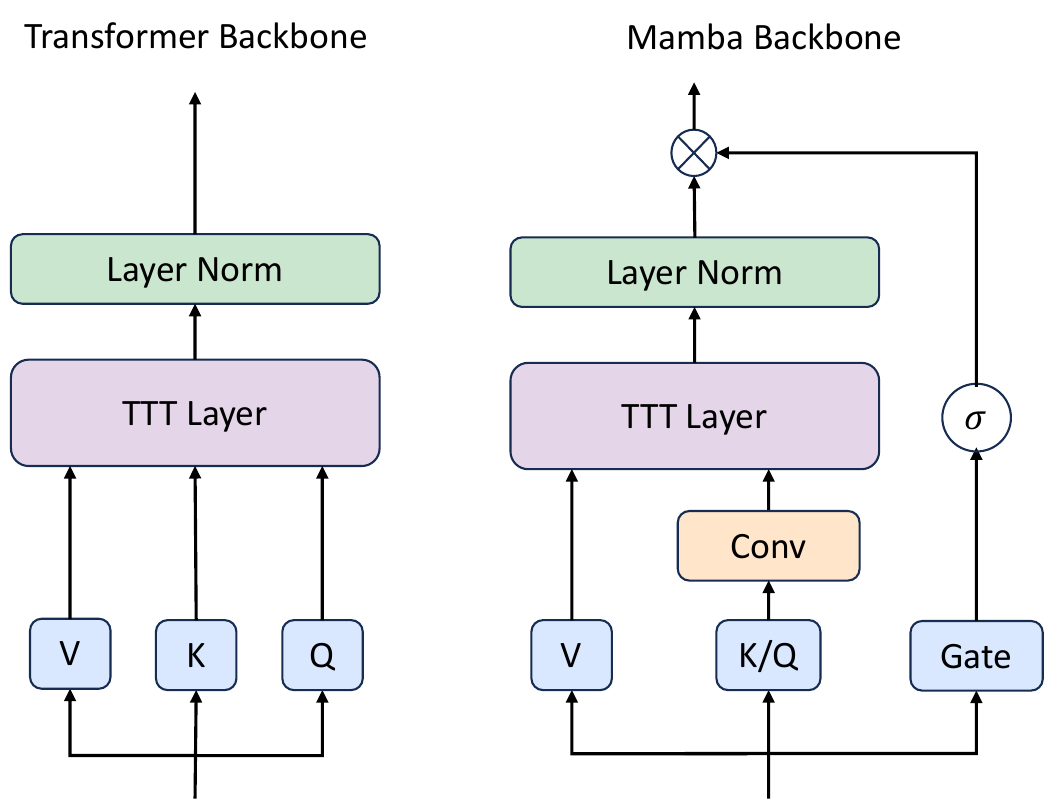}
  \caption{Structure of the Sequence Modeling Block. In the Mamba backbone, $\sigma$ represents GELU \cite{hendrycks2016gaussian}.}
  \label{fig_bone}
\end{figure}

As shown in Fig.~\ref{fig_bone}, both the Transformer and Mamba backbones incorporate the \textbf{TTT layer}, where a gradient-based update of the self-supervised inner loop is applied to adjust the model’s parameters during test time. Besides, $\theta_V$, $\theta_K$, and $\theta_Q$ are learnable matrices that optimized in the outer loop. Inspired by architectures like NormFormer \cite{shleifer2021normformer}, the two variants also include an additional layer normalization step before the output. In the Mamba backbone, the TTT layer incorporates a gated activation mechanism using GELU \cite{hendrycks2016gaussian}. This allows the model to keep and forget long sequence information efficiently. To accommodate the additional gating parameters, we merge the $\theta_K$ and $\theta_Q$ projections into a single projection, reducing the dimensionality without sacrificing expressive power. Meanwhile, before fed to the TTT layer, $\theta_K$ and $\theta_Q$ are processed by a 1d-convolution layer to capture local relationships from the input sequence.

Finally, after the sequence modeling block, a \textbf{feed-forward block} is applied, which consists of two fully connected layers with a non-linear activation in between. The residual connection then ensures that the output of the block can bypass the sequence modeling and feed-forward transformations, helping to mitigate vanishing gradients and enabling deeper architectures to propagate information effectively.

\subsection{Training Process of the TTT-based Sequence Modeling Block}

The sequence modeling block in TTT4Rec leverages a two-loop training process, with the \textbf{outer loop} handling the global supervised learning task, and the \textbf{inner loop} focusing on self-supervised updates during both training and inference. 

\subsubsection{Outer Loop (Supervised Learning)}

The outer loop is responsible for the supervised learning process during training. It updates the global model parameters \(\theta_K\), \(\theta_Q\), and \(\theta_V\), which are used to generate different projections of the input sequence for the self-supervised learning tasks. The supervised loss function \(\mathcal{L}_{\text{outer}}\) is typically defined as the cross-entropy loss for next-item prediction:

\[
\mathcal{L}_{\text{outer}} = - \log P(v_{n_u+1} | v_1, \dots, v_{n_u}; \theta),
\]
where \(\theta\) represents the global model parameters, and \(P(v_{n_u+1} | v_1, \dots, v_{n_u}; \theta)\) is the probability of predicting the next item \(v_{n_u+1}\) based on the current sequence. The parameters \(\theta_K\), \(\theta_Q\), and \(\theta_V\) are updated via gradient descent:

\[
\theta_K, \theta_Q, \theta_V \leftarrow \theta_K, \theta_Q, \theta_V - \eta_{\text{outer}} \nabla_{\theta} \mathcal{L}_{\text{outer}},
\]
where \(\eta_{\text{outer}}\) is the learning rate for the outer loop. These global parameters define different projections of the input sequence used for the inner loop's self-supervised learning.

\subsubsection{Inner Loop (Self-Supervised Learning)}

The inner loop is responsible for updating the model's hidden state \(W_t\) during both training and test time. It performs gradient-based updates using a self-supervised loss function \(\mathcal{L}_{\text{inner}}\), which allows the model to adapt its parameters dynamically as new user interactions are observed.

In the inner loop, each input token \(x_t\) is projected into different views using the global parameters \(\theta_K\), \(\theta_Q\), and \(\theta_V\). These projections are used to compute the self-supervised loss and update the hidden state \(W_t\). Specifically, the input \(x_t\) is transformed into a \textit{training view} and a \textit{label view} as follows:

\[
\text{Training view: } x_{t}^{(K)} = \theta_K x_t, \quad \text{Label view: } x_{t}^{(V)} = \theta_V x_t.
\]

The model then attempts to predict the label view from the training view using the hidden state \(W_t\). The self-supervised loss is defined as the reconstruction error between these views:

\[
\mathcal{L}_{\text{inner}} = \| f(x_t^{(K)}; W_{t-1}) - x_t^{(V)} \|^2,
\]
where \(f(x_t^{(K)}; W_{t-1})\) is the model's prediction of \(x_t^{(V)}\) using the current hidden state \(W_{t-1}\). The prediction model $f(\cdot;W)$ could be a linear function or an MLP. The hidden state is updated through gradient descent based on the self-supervised loss:

\[
W_t = W_{t-1} - \eta_{\text{inner}} \nabla_{W_{t-1}} \mathcal{L}_{\text{inner}},
\]
where \(\eta_{\text{inner}}\) is the learning rate for the inner loop. This process continues throughout the sequence, allowing the model to adjust its hidden state dynamically.

The inner loop performs these updates both during training and at test time, enabling the model to refine its understanding of user behavior as new interactions are observed. This continuous adaptation is crucial for handling dynamic user behaviors and improving the model's prediction accuracy during inference.

\subsubsection{Multi-View Projections and Output Generation}

In addition to the training and label views used for the inner loop, the model also generates a \textit{test view} to produce the final output for the next-item prediction. The test view is computed using the global parameter \(\theta_Q\):

\[
\text{Test view: } x_{t}^{(Q)} = \theta_Q x_t.
\]

The final output of the TTT layer is then generated by applying the updated hidden state \(W_t\) to the test view:

\[
o_{t+1} = f(x_{t}^{(Q)}; W_t).
\]

\subsubsection{Combined Optimization}

The training process alternates between the outer loop and the inner loop, ensuring that both global parameters \(\theta_K\), \(\theta_Q\), \(\theta_V\) and the hidden state \(W_t\) are optimized. The outer loop updates the global parameters using supervised learning, while the inner loop refines the hidden state during both training and inference via self-supervised learning. This dual-loop optimization ensures that TTT4Rec can adapt to new user interactions and dynamically improve its recommendation performance.

By combining the outer-loop supervised learning with inner-loop self-supervised updates, TTT4Rec can handle dynamic user behaviors more effectively than traditional static models. This allows the model to continuously refine its understanding of user preferences in real-time, resulting in more accurate and personalized recommendations.

\subsection{Prediction Layer}

In TTT4Rec, the Prediction Layer is responsible for generating the final recommendation based on the sequence of embeddings produced by the previous layers. We use the embedding of the last item in the sequence to predict the next item a user is likely to interact with.

Formally, given the sequence of item embeddings \( \hat{H} = \{\hat{h}_1, \hat{h}_2, \dots, \hat{h}_{n_u}\} \) produced by residual blocks, where \( \hat{h}_{n_u} \) represents the embedding of the last item \( v_{n_u} \) in the user’s interaction history, the prediction layer computes the probability distribution over all possible items in the item set \( \mathcal{V} \). This is done by projecting \( \hat{h}_{n_u} \) into the item space and applying a softmax function:

\[
\hat{y} = \text{Softmax}(\hat{h}_{n_u} M^{\top}),
\]
where \( M \in \mathbb{R}^{|\mathcal{V}| \times D} \) is the learnable weight matrix that maps the final item embedding \( \hat{h}_{n_u} \) to the probability space over the item set \( \mathcal{V} \), and \( \hat{y} \in \mathbb{R}^{|\mathcal{V}|} \) represents the probability distribution for the next item.

The item with the highest probability is selected as the predicted next interaction:

\[
\hat{v}_{n_u+1} = \arg\max_{v \in \mathcal{V}} \hat{y}_v.
\]

\section{Experiments}

\subsection{Experimental Setup}

In this section, we evaluate the performance of TTT4Rec on three widely-used datasets for sequential recommendation. These datasets cover a range of application domains, including location-based check-ins, live streaming recommendations, and e-commerce product purchases.

\subsubsection{Datasets}

We use the following three datasets in our experiments:

\begin{itemize}
    \item \textbf{Gowalla} \cite{liu2013personalized}: This dataset comes from a location-based social networking platform where users share their locations by checking in at various venues. It contains check-ins made by users from February 2009 to October 2010.
    
    \item \textbf{Twitch-100k}: This dataset captures user interactions with streaming content on the Twitch platform. It contains data collected over a 43-day period.
    
    \item \textbf{Amazon-video-game} \cite{mcauley2015image}: This dataset includes product reviews and ratings for video games from Amazon, allowing us to model user preferences in the e-commerce domain.
\end{itemize}

Table \ref{tab:dataset_stats} provides statistics for the datasets, including the number of users, items, interactions, and the average sequence length.

\begin{table}[htpb]
    \small
    \centering
    \begin{tabular}{lcccc}
        \toprule
        \multirow{2}{*}{\textbf{Dataset}} & \multirow{2}{*}{\textbf{\# Users}} & \multirow{2}{*}{\textbf{\# Items}} & \multirow{2}{*}{\textbf{\# Inter.}} & \textbf{Avg. Seq.} \\
        &          &             &                     & \textbf{Length} \\
        \midrule
        Gowalla              & 107,092 & 1,280,969& 6,442,892 & 37.18 \\
        Twitch-100k          & 100,000 & 739,991  & 3,051,732 & 30.52 \\
        Amazon-video-game    & 826,767  & 50,210  & 1,324,753 & 9.54 \\
        \bottomrule
    \end{tabular}
    \caption{Statistics of the datasets used in the experiments.}
    \label{tab:dataset_stats}
\end{table}

\subsubsection{Baselines}

We compare TTT4Rec against several competitive baseline models that have been widely used in the sequential recommendation literature. These include:

\begin{itemize}
    \item \textbf{GRU4Rec} \cite{tan2016improved}: An RNN-based model that leverages gated recurrent units (GRU) for session-based recommendation.
    \item \textbf{NARM} \cite{li2017neural}: A neural attentive model for session-based recommendation that incorporates an attention mechanism to capture user behavior.
    \item \textbf{SASRec} \cite{kang2018self}: A Transformer-based sequential recommendation model that uses self-attention mechanisms to capture long-term dependencies in user behavior.
    \item \textbf{BERT4Rec} \cite{sun2019bert4rec}: A sequential recommendation model based on the BERT architecture, leveraging bidirectional self-attention.
    \item \textbf{Mamba4Rec} \cite{liu2024mamba4rec}: A state-space model (SSM) that efficiently models long-range dependencies using an SSM-based architecture.
\end{itemize}

\subsubsection{Evaluation Metrics}

To assess the performance of TTT4Rec and the baseline models, we adopt the following commonly used evaluation metrics in sequential recommendation:

\begin{itemize}
    \item \textbf{Hit Ratio (HR)}: Measures whether the ground truth item appears in the top-K recommended items. We report HR@10 and HR@50.
    \item \textbf{Normalized Discounted Cumulative Gain (NDCG)}: Measures the ranking quality by assigning higher scores to correctly recommended items that appear higher in the list. We report NDCG@10 and NDCG@50.
\end{itemize}

\subsubsection{Implementation Details}

\begin{table*}[htpb]
    \centering
    \setlength{\tabcolsep}{3.5pt}
    \begin{tabularx}{\textwidth}{p{1.7cm} | >  {\centering\arraybackslash}X > {\centering\arraybackslash}X > {\centering\arraybackslash}X > {\centering\arraybackslash}X | > {\centering\arraybackslash}X > {\centering\arraybackslash}X > {\centering\arraybackslash}X >{\centering\arraybackslash}X | >  {\centering\arraybackslash}X >{\centering\arraybackslash}X >{\centering\arraybackslash}X >{\centering\arraybackslash}X}
        \toprule
        \multirow{2}{*}{\textbf{Models}} & \multicolumn{4}{c}{\textbf{Gowalla}} & \multicolumn{4}{c}{\textbf{Twitch-100k}} & \multicolumn{4}{c}{\textbf{Amazon-video-game}} \\
        \cmidrule(lr){2-5}  \cmidrule(lr){6-9}  \cmidrule(lr){10-13} 
        & \footnotesize \textbf{HR@10} & \footnotesize \textbf{HR@50} & \footnotesize \textbf{NDCG@10} & \footnotesize \textbf{NDCG@50} & \footnotesize \textbf{HR@10} & \footnotesize \textbf{HR@50} & \footnotesize \textbf{NDCG@10} & \footnotesize \textbf{NDCG@50} & \footnotesize \textbf{HR@10} & \footnotesize \textbf{HR@50} & \footnotesize \textbf{NDCG@10} & \footnotesize \textbf{NDCG@50} \\
        \midrule
        \textbf{GRU4Rec}  & 0.0306 & 0.0840 & 0.0156 & 0.0271 & 0.0391 & 0.0917 & 0.0206 & 0.0320 & 0.0495 & 0.1428 & 0.0246 & 0.0446 \\
        \textbf{NARM}     & 0.0332 & 0.0896 & 0.0172 & 0.0293 & 0.0382 & 0.0830 & 0.0208 & 0.0306 & 0.0544 & 0.1590 & 0.0268 & 0.0494 \\
        \textbf{SASRec}   & \textbf{0.0934} & \textbf{0.1933} & \underline{0.0425} & \underline{0.0644} & \underline{0.0601} & \textbf{0.1339} & 0.0276 & \underline{0.0439} & \underline{0.0841} & \underline{0.2107} & \underline{0.0401} & \underline{0.0677} \\
        \textbf{BERT4Rec} & 0.0363 & 0.0994 & 0.0181 & 0.0317 & 0.0201 & 0.0592 & 0.0099 & 0.0183 & 0.0259 & 0.0845 & 0.0126 & 0.0251 \\
        \textbf{Mamba4Rec}& 0.0572 & 0.1179 & 0.0339 & 0.0471 & 0.0525 & 0.1124 & \underline{0.0292} & 0.0423 & 0.0711 & 0.1732 & 0.0394 & 0.0615 \\
        \midrule
        \textbf{TTT4Rec}  & \textbf{0.0934} & \underline{0.1893} & \textbf{0.0449} & \textbf{0.0659} & \textbf{0.0623} & \underline{0.1299} & \textbf{0.0309} & \textbf{0.0458} & \textbf{0.0879} & \textbf{0.2178} & \textbf{0.0425} & \textbf{0.0707} \\
        Improv. & 0\% & - & 5.65\% & 2.33\% & 3.66\% & - & 5.82\% & 4.33\% & 4.52\% & 3.37\% & 5.99\% & 4.43\% \\
        \bottomrule
    \end{tabularx}
    \caption{Performance comparison. The best results are in bold, and the second-best are underlined. The improvement percentage is calculated between TTT4Rec and the best baseline model.}
    \label{tab:overall_performance}
\end{table*}

In the default architecture of TTT4Rec, we employ a single residual block with the transformer backbone. A two-layer MLP with GELU \cite{hendrycks2016gaussian} activation is used as the inner-loop prediction function. Rotation factor \( \mu \) is set to be 1000. For all models, we adopt the Adam optimizer \cite{kingma2014adam} with a learning rate of 0.001. The training batch size is set to 2048, while the evaluation batch size is 4096. We use an embedding dimension of 64 and a hidden state dimension of 256.

To evaluate the model’s performance under conditions of limited data and high variability in user interests, we split each dataset into training, validation, and test sets in a 3:2:5 ratio. This split is done in chronological order based on each user’s interaction history. The training set contains fewer interactions (representing the limited data condition), while the test set includes longer sequences (allowing for more potential variations in user interests). To ensure the integrity of user interaction sequences under this split, we define a minimum sequence length for each dataset, which is proportional to the average interaction length in the corresponding dataset. Specifically, the minimum sequence length for Gowalla is 20, for Twitch-100k it is set to 15, and for Amazon-video-game it is 5.

The maximum context length considered by models is also set according to the average interaction length in each dataset. For Gowalla, the maximum context length is set to 100, for Twitch-100k it is 70, and for Amazon-video-game it is 50. This ensures that the model captures sufficient historical context while aligning with the interaction patterns of each dataset.

For further implementation details, we follow the configurations provided by RecBole \cite{zhao2021recbole}, a unified recommendation benchmark that ensures consistent experimental settings across all models.

\subsection{Overall Performance}

In this section, we compare the performance of TTT4Rec with five baseline models across the three datasets. The results are shown in Table \ref{tab:overall_performance}. TTT4Rec consistently outperforms or matches the state-of-the-art baseline models. On the Gowalla dataset, TTT4Rec achieves the best performance in NDCG@10 and NDCG@50, tying with SASRec in HR@10 and surpassing all other models in NDCG metrics. Similarly, on Twitch-100k, TTT4Rec surpasses all baselines, obtaining the highest scores for HR@10, NDCG@10, and NDCG@50. For the Amazon-video-game dataset, TTT4Rec shows significant superiority, achieving the best results across all metrics. These results demonstrate that TTT4Rec consistently outperforms RNN-based models, attention-based models, and the state-space model, particularly in HR@10 and NDCGs, which measure both ranking and hit accuracy.

\subsection{Evaluation of TTT4Rec Variants}
In this section, we evaluate the performance of four different TTT4Rec variants on the Gowalla and Amazon-video-game dataset. The evaluation compares two sequence modeling backbones, Transformer and Mamba, paired with two inner-loop prediction models: Linear and MLP. 

\begin{table}[htpb]
    \centering
    \setlength{\tabcolsep}{6pt}
    \begin{tabularx}{\linewidth}{p{1.4cm} p{1.3cm} >{\centering\arraybackslash}X >{\centering\arraybackslash}X >{\centering\arraybackslash}X >{\centering\arraybackslash}X}
        \toprule
        &  & \multicolumn{4}{c}{\textbf{TTT4Rec Variants}} \\
        \cmidrule(lr){3-6}
        \multirow{2}{*}{\textbf{Dataset}} & \multirow{2}{*}{\textbf{Metrics}} & \makecell{ \small \textbf{Mamba+} \\ \small \textbf{Linear}} & \makecell{\small \textbf{Mamba+} \\ \small \textbf{MLP}} & \makecell{\small \textbf{Trans+} \\ \small \textbf{Linear}} & \makecell{\small \textbf{Trans+} \\ \small \textbf{MLP}} \\
        \midrule
        \multirow{4}{*}{Gowalla} 
        & HR@10     & 0.0845 & 0.0845 & \underline{0.0888} & \textbf{0.0934} \\
        & HR@50     & 0.1806 & 0.1784 & \underline{0.1849} & \textbf{0.1893} \\
        & NDCG@10   & 0.0412 & 0.0409 & \underline{0.0421} & \textbf{0.0449} \\
        & NDCG@50   & 0.0621 & 0.0614 & \underline{0.0631} & \textbf{0.0659} \\
        \midrule
        \multirow{4}{*}{\makecell[l]{Amazon- \\ video-game}} 
        & HR@10     & 0.0861 & 0.0800 & \textbf{0.0899} & \underline{0.0879} \\
        & HR@50     & 0.2158 & 0.2162 & \textbf{0.2232} & \underline{0.2178} \\
        & NDCG@10   & 0.0421 & 0.0418 & \textbf{0.0440} & \underline{0.0425} \\
        & NDCG@50   & 0.0703 & 0.0703 & \textbf{0.0730} & \underline{0.0707} \\
        \bottomrule
    \end{tabularx}
    \caption{Performance comparison of TTT4Rec variants on the Gowalla and Amazon-video-game datasets. The best results are bold, and the second-best are underlined.}
    \label{tab:ttt4rec_variants}
\end{table}

The experimental results shown in Table ~\ref{tab:ttt4rec_variants} indicate that the TTT4Rec variant with the Transformer backbone consistently outperforms the Mamba backbone across both the Gowalla and Amazon-video-game datasets. This may suggest that the Transformer backbone is better suited for sequential recommendation tasks. However, the underlying mechanisms contributing to this superior performance still require further investigation.

On the Gowalla dataset, the Transformer+MLP variant achieves the best performance across all metrics. Interestingly, the Amazon-video-game dataset shows a different pattern, where the Transformer+Linear variant performs best, and the Transformer+MLP variant ranks second. MLP performs well on the Gowalla dataset, where the item sequence is longer, suggesting that the MLP’s additional flexibility is beneficial in capturing user preferences in more intricate datasets. In contrast, the Linear model shows a stronger performance on the smaller Amazon-video-game dataset, likely due to its simplicity and lower risk of overfitting.

\subsection{Experiments under Different Data Split Ratios}

In this section, we evaluate the performance of TTT4Rec and baseline models on the Amazon-video-game dataset, using a training/validation/test split ratio of 6:2:2 rather than 3:2:5. This setup simulates a scenario with more training data, allowing us to analyze the models' performance under conditions where the model has more learning opportunities.

\begin{table}[htpb]
    \centering
    \setlength{\tabcolsep}{4pt}  
    \begin{tabularx}{\linewidth}{p{1.7cm} >{\centering\arraybackslash}X >{\centering\arraybackslash}X >{\centering\arraybackslash}X >{\centering\arraybackslash}X}
        \toprule
        \multirow{2}{*}{\textbf{Models}} & \multicolumn{4}{c}{\textbf{Amazon-video-game}} \\
        \cmidrule(lr){2-5}
        & HR@10 & HR@50 & NDCG@10 & NDCG@50 \\
        \midrule
        \textbf{GRU4Rec}   & 0.0764 & 0.2070 & 0.0389 & 0.0671 \\
        \textbf{NARM}      & 0.0879 & 0.2297 & 0.0449 & 0.0755 \\
        \textbf{SASRec}    & \underline{0.1016} & \underline{0.2546} &  0.0482 & \underline{0.0814} \\
        \textbf{BERT4Rec}  & 0.0580 & 0.1691 &  0.0292 & 0.0531 \\
        \textbf{Mamba4Rec} & 0.0931 & 0.2309 & \underline{0.0504} & 0.0803 \\
        \midrule
        \textbf{TTT4Rec}   & \textbf{0.1034} & \textbf{0.2592} & \textbf{0.0509} & \textbf{0.0848} \\
        Improv.  & 1.77\% & 1.81\% & 1.00\% & 4.18\% \\
        \bottomrule
    \end{tabularx}
    \caption{Performance comparison of TTT4Rec and baselines on Amazon-video-game dataset with a 6:2:2 data split ratio. Best results are in bold, and the second-best are underlined. The improvement percentage is calculated between TTT4Rec and the best baseline model.}
    \label{tab:ttt4rec_622_amazon}
\end{table}

As shown in Table \ref{tab:ttt4rec_622_amazon}, TTT4Rec continues to outperform all baseline models across all metrics on the Amazon-video-game dataset with a 6:2:2 training/validation/test split. However, compared to the second-best performing model, the performance improvements are smaller than those observed under the 3:2:5 split. This implies that TTT4Rec’s benefits are more pronounced in scenarios where the training data is limited, or when user behavior is more volatile during inference, making it particularly well-suited for such conditions.

\section{Conclusion}

In this paper, we introduced TTT4Rec, a novel approach for sequential recommendation that leverages the Test-Time Training (TTT) paradigm to continuously update model parameters using unsupervised learning during the test phase. Our model demonstrated strong performance across multiple sequential recommendation datasets, including Gowalla, Twitch-100k, and Amazon-video-game. Notably, TTT4Rec consistently outperformed state-of-the-art baselines such as GRU4Rec, NARM, SASRec, BERT4Rec, and Mamba4Rec.

Through extensive experiments, we observed that TTT4Rec shows a significant advantage in scenarios where training data is limited or user behaviors are volatile, such as with the 3:2:5 data split ratio. Under these conditions, TTT4Rec’s ability to adapt at test time allows it to model user preferences more effectively than traditional methods. Moreover, our ablation study on different TTT4Rec variants revealed that the Transformer backbone consistently outperforms the Mamba backbone, though further investigation is needed to fully understand the mechanisms driving this superiority. Additionally, simpler inner-loop models like Linear proved effective on smaller datasets, such as Amazon-video-game, possibly due to their ability to avoid overfitting.

In conclusion, TTT4Rec provides a flexible and powerful solution for sequential recommendation tasks, particularly in data-limited or high-variance environments. Future work will explore the underlying mechanisms of TTT and further optimize the model for real-world applications with fluctuating user behaviors and sparse data.


\bibliographystyle{ACM-Reference-Format}
\bibliography{references.bib}

\end{document}